\documentclass[9pt,twocolumn,twoside]{osajnl}

\journal{optica} 

\setboolean{shortarticle}{true}

\title{Ultra-broadband quadrature squeezing with thin-film lithium niobate nanophotonics}

\author[1,2]{Pao-Kang Chen}
\author[1]{Ian Briggs}
\author[1]{Songyan Hou}
\author[1,*]{Linran Fan}
\affil[1]{J. C. Wyant College of Optical Sciences, University of Arizona, 1630 E. University Boulevard, Tucson, Arizona 85721, USA}
\affil[2]{Department of Physics, University of Arizona, 1118 E. Fourth Street, Tucson, Arizona 85721, USA}
\affil[*]{Corresponding author: lfan@optics.arizona.edu}

\begin{abstract}
Squeezed light is a key quantum resource that enables quantum advantages for sensing, networking, and computing applications. The scalable generation and manipulation of squeezed light with integrated platforms are highly desired for the development of quantum technology with continuous variables. In this letter, we demonstrate squeezed light generation with thin-film lithium niobate integrated photonics. Parametric down-conversion is realized with quasi-phase matching using ferroelectric domain engineering. With sub-wavelength mode confinement, efficient nonlinear processes can be observed with single-pass configuration. We measure $0.56\pm0.09$\;dB quadrature squeezing ($\sim$3\;dB inferred on-chip). The single-pass configuration further enables the generation of squeezed light with large spectral bandwidth up to 7\;THz. This work represents a significant step towards the on-chip implementation of continuous-variable quantum information processing.
\end{abstract}

\setboolean{displaycopyright}{true}

\begin{document}

\maketitle

Squeezed light is a quintessential quantum resource with no classical counterpart. With the recent development of quantum information, squeezed light has become one fundamental building block for photonic quantum technology \cite{weedbrook2012gaussian,braunstein2005quantum}. The suppression of measurement variance below shot noise has been widely recognized and implemented for sensing applications including gravitational wave detection \cite{ligo1_Abadie2011}, microscopy \cite{microscopy1_Ono2013}, and bio-sensing \cite{bio-sensing1_Taylor2013}. Furthermore, squeezed light also plays a critical role in quantum communications, with demonstrations of unconditional quantum teleportation \cite{quantum_Teleportation1_Furusawa706} and quantum key distribution \cite{madsen2012continuous}. Continuous-variable quantum computing has also been developed based on squeezed light \cite{quantum_computing1_PhysRevA.79.062318,menicucci2014fault}. It provides an important alternative with unique features to the discrete-variable approach using single photons. Quantum information is encoded into the continuous quadrature amplitudes instead of discrete degrees of freedom (such as polarization). This renders the deterministic generation and entanglement operation of quantum states \cite{quantum_computing2_Yokoyama2013,chen2014experimental}. Deterministic, universal, and fault-tolerant quantum computation can be realized \cite{quantum_computing1_PhysRevA.79.062318,menicucci2014fault}.

The further development of continuous-variable quantum technology requires a large number of squeezed light sources and complex photonic circuits \cite{quantum_computing2_Yokoyama2013,chen2014experimental,guo2020distributed}. Integrated photonics provides the scalable fabrication of quantum sources, with additional advantages including phase stability, near-perfect spatial mode matching, and power efficiency \cite{elshaari2020author}. Currently, squeezed light has been generated using four-wave mixing in integrated photonic ring cavities \cite{nano_ring1_Arrazola2021, nano_ring2_Vaidyaeaba9186, nano_ring3_yang2021squeezed, nano_ring4_PhysRevApplied.3.044005, nano_ring5_PhysRevLett.124.193601}. The small wavelength separation between pump and squeezed light makes it challenging to remove the pump without introducing extra loss on squeezed light. The pump efficiency is low compared with parametric down-conversion, which is more widely used for squeezed light generation in free-space experiments \cite{andersen201630}. These platforms also lack electro-optic effect. Therefore, thermo-optic or electro-optomechanical effect are used for phase shifting, which limit the speed of photonic circuit reconfiguration \cite{elshaari2020author}. This prevents the implementation of critical functions such as feedforward measurement, which is essential for measurement-based one-way quantum computing\;\cite{mbqc1_Prevedel2007,mbqc2_Walther2005}. Squeezed light has also been generated in bulk lithium niobate waveguide \cite{single_pass1_doi:10.1063/1.5142437, single_pass_LN1_Lenzinieaat9331, single_pass_LN2_Mondain:19}. Parametric down-conversion and fast reconfiguration can be achieved. However, the low index contrast and large cross-sections prevent the scalable integration of large-scale photonic circuits.

\begin{figure}[t]
\centering
\includegraphics[width=\linewidth]{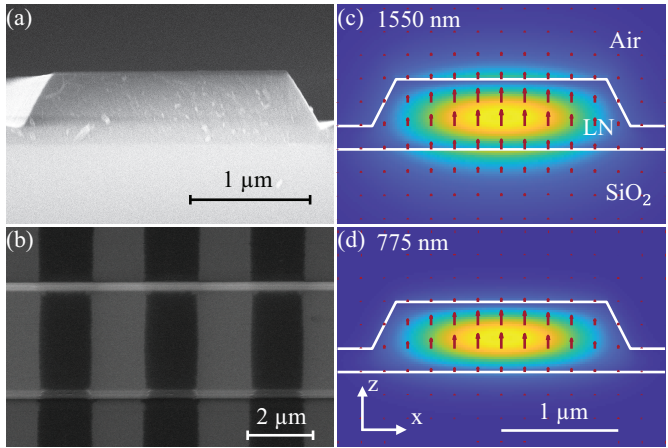}
\caption{Design and fabrication of squeezed light sources on thin-film lithium niobate. (a) Scanning electron microscope image of the fabricated device cross-section. (b) Scanning electron microscope top-view image of the mock-up device etched in hydrofluoric acid. (c) Simulated mode profile and electric field of the squeezed light mode at 1550\;nm. (d) Simulated mode profile and electric field of the pump mode at 775\;nm.}
\label{figure1}
\end{figure}

In this letter, we report the development of thin-film lithium niobate integrated photonics for squeezed light generation. The strong second-order nonlinearity and sub-wavelength confinement of optical fields facilitate efficient parametric down-conversion processes in a single-pass configuration. This eliminates the need of photonic cavities for pump recycling. Therefore, squeezed light can escape the device with near-unity efficiency, in contrast to the cavity configuration where the challenging highly over-coupled condition is required \cite{andersen201630}. Quasi-phase matching is realized through periodic poling of thin-film lithium niobate. We have measured $0.56\pm0.09$\;dB continuous-wave quadrature squeezing, with inferred $\sim$3\;dB on-chip squeezing. Our device further achieves squeezing over 7\;THz bandwidth, covering S-, C-, and L-bands for optical communications.

\begin{figure}[t]
\centering
\includegraphics[width=\linewidth]{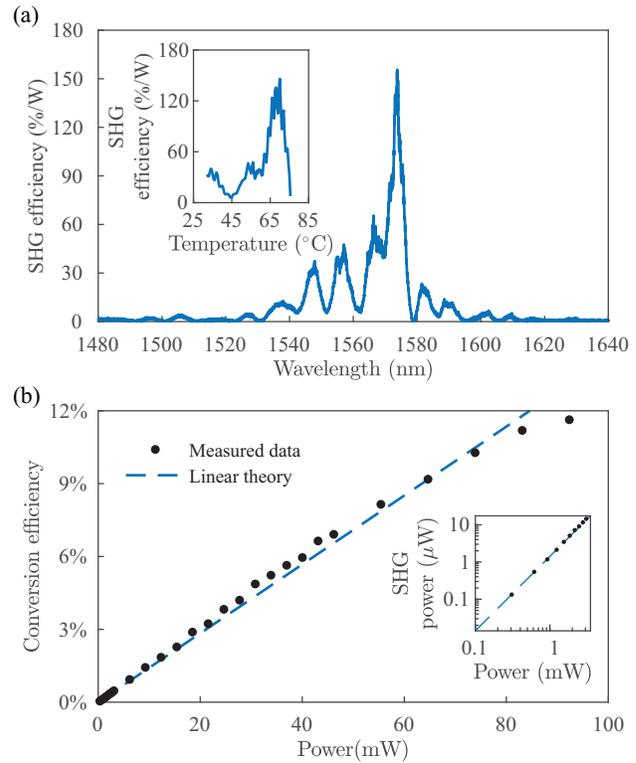}
\caption{Second harmonic generation with periodically poled thin-film lithium niobate waveguide. (a) Measured SHG efficiency versus input wavelength at room temperature with input power 2\;mW. Inset: Measured SHG efficiency versus device temperature with fixed input wavelength 1554.2\;nm. (b) Measured SHG efficiency versus input power. Experimental data (black dot) are plotted with a linear fit of the low power region (blue dashed line). The inset shows the SHG signal at low input power.}
\label{figure2}
\end{figure}

The device is fabricated from a z-cut thin-film lithium niobate wafer with 600\;nm device layer (Fig.\;\ref{figure1}(a)). The top width of the ridge waveguide is 1.8\;\textmu m, defined by electron-beam lithography with hydrogen silsesquioxane resist. The pattern is transferred to the lithium niobate device layer using argon-based plasma processing with 400\;nm etching depth. We use the fundamental transverse-magnetic (TM$_{\rm 00}$) modes for both the squeezed light around wavelength 1550\;nm (Fig.\;\ref{figure1}(c)) and the pump around wavelength 775\;nm (Fig.\;\ref{figure1}(d)) to utilize the largest second-order nonlinear component $d_{33}=-25$\;pm/V in lithium niobate. In order to compensate for the wavevector mismatch between the squeezed light $k_s$ and the pump $k_p$, an additional wavevector is introduced by periodically inverting the crystal polarization of the thin-film lithium niobate. The quasi-phase matching condition is fulfilled when the domain period $\Lambda=2\pi/(2k_s-k_p)$\;\cite{1650535}. In order to realize the periodic domain inversion, nickel electrodes are patterned on top of the ridge waveguide with period $\Lambda \approx 3$\;\textmu m. Multiple high-voltage pulses are applied to nickel electrodes at an elevated temperature. Then nickel electrodes are removed by hydrochloric acid. Figure\;\ref{figure1}(b) shows the mock-up device after etching in hydrofluoric acid \cite{HFetch_B106279B}. A duty cycle close to 50\% can be clearly observed. The total length of the device with domain inversion is 5\;mm. Finally, the chip is cleaved to expose the waveguide facets for input and output coupling.

\begin{figure*}[t]
\centering
\includegraphics[width=\textwidth]{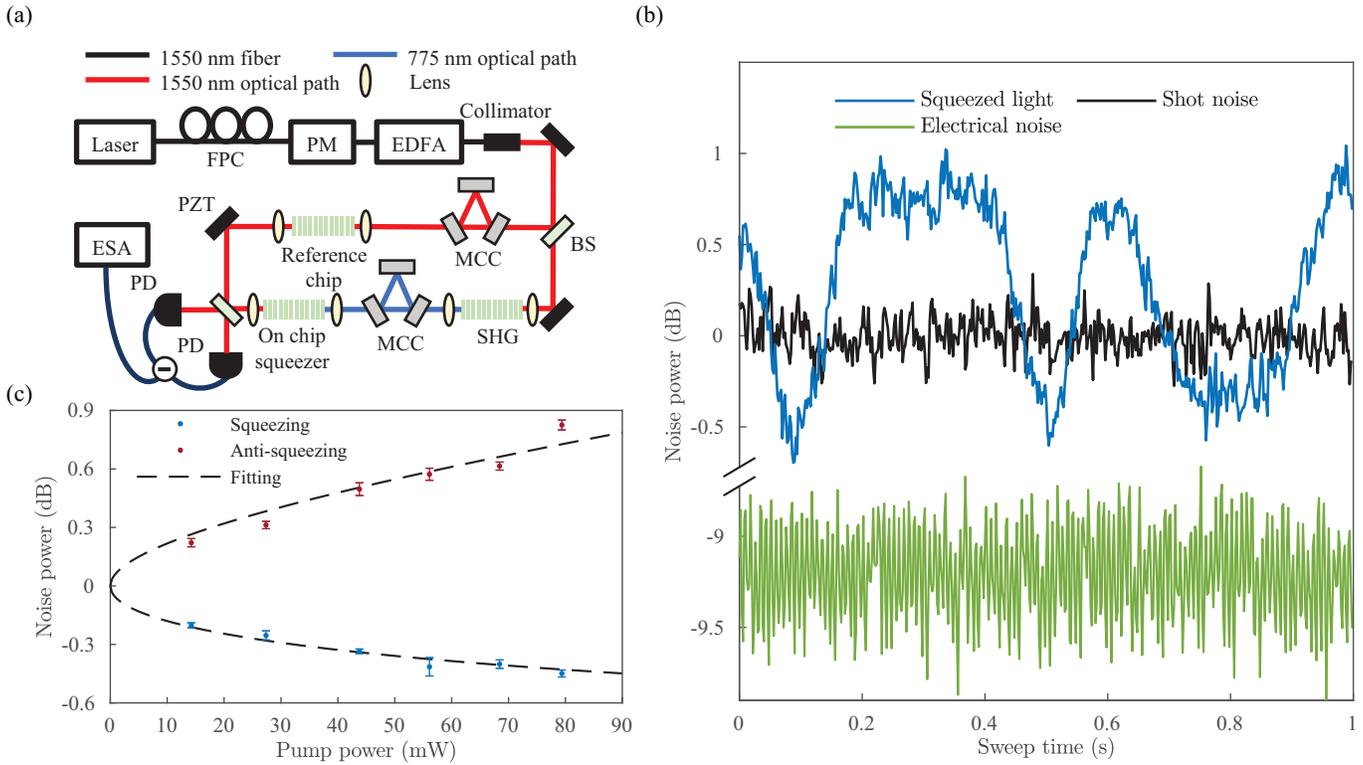}
\caption{Squeezed light measurement. (a) Experimental setup. FPC: fiber polarization controller; PM: optical phase modulator; EDFA: erbium-doped fiber amplifier; BS: 50/50 beamsplitter; MCC: mode cleaning cavity; PD: photodetector; ESA: electrical spectrum analyzer; PZT: piezoelectric transducer. (b) Normalized noise power as a function of time with phase sweeping. Data obtained with zero span mode at frequency 22\;MHz. The resolution bandwidth is 300 KHz. The video bandwidth is 100 Hz. (c) Squeezing and anti-squeezing levels as a function of pump power at measurement frequency 22 MHz.}
\label{figure3}
\end{figure*}

We use second harmonic generation (SHG) to determine the phase-matching condition and evaluate the efficiency of the nonlinear optical process in the fabricated device. For SHG measurement, we use optical fields near 1550\;nm as input, and detect optical output power near 775\;nm (Fig.\;\ref{figure2}(a)). At room temperature, the device has the maximum SHG output power around 1574\;nm. The SHG efficiency 155\%/W is estimated. By tuning the device temperature, the phase-matching condition can be adjusted. At temperature 70\;$^{\circ}C$, the optimum wavelength is shifted to 1554.2\;nm without compromising the SHG efficiency (Fig.\;\ref{figure2}(a) inset). At low power levels, SHG power conversion efficiency increases linearly with the input power. At high power levels, the conversion efficiency deviates from this behavior due to pump depletion, which is observed with our device (Fig.\;\ref{figure2}(b)). We have achieved more than 10\% total conversion efficiency, showing that efficient nonlinear processes can be realized with single-pass configuration.

The experimental setup for squeezing measurement is shown in Fig.\;\ref{figure3}(a). A continuous-wave (CW) laser with center wavelength 1554.2\;nm is phase modulated at 40\;MHz, and amplified by an erbium-doped fiber amplifier (EDFA). The EDFA output is collimated and divided into two beams. One beam is injected into a commercial nonlinear crystal for SHG. The second-harmonic beam is used as the pump for the squeezed light generation with thin-film lithium niobate waveguides. The other beam is used as the local oscillator (LO) for homodyne detection. The LO beam passes through a reference waveguide with identical dimensions as the thin-film lithium niobate device for squeezed light generation. This ensures the optimum spatial mode matching between the LO and squeezed light. Mode cleaning cavities, which are locked to the CW laser through Pound–Drever–Hall (PDH) technique using the 40\;MHz modulation signal, are used to remove excess noise in the pump field and LO. The homodyne signal is sent to an electrical spectrum analyzer (ESA) for noise measurement. A piezo-transducer is used to scan the LO phase with 1\;Hz frequency. The device temperature is tuned to maximize the nonlinear efficiency at 1554.2\;nm (Fig.\;\ref{figure2}(a) inset).

Figure\;\ref{figure3}(b) shows the noise power of the squeezed light generated with the thin-film lithium niobate waveguide. The noise power is detected by zero span measurement at 22\;MHz frequency. With 82\;mW pump power, the measured squeezing and anti-squeezing levels are $-0.56\pm0.09$\;dB and $0.83\pm0.09$\;dB, respectively. The shot noise level is determined by blocking the output of the thin-film lithium niobate waveguide. With 12\;mW LO power, the shot noise is 9.2\;dB higher than the electronic noise of the measurement setup. We further measure the squeezing and anti-squeezing levels with respect to pump power (Fig.\;\ref{figure3}(c)). The measured data are fitted with
\begin{equation}
S_{\pm}= 1-T+Te^{\pm 2\sqrt{\eta P}}
\end{equation}
where $S_{\rm \pm}$ are the anti-squeezing and squeezing levels in linear scale; $T$ is the total detection efficiency; $\eta$ is the SHG efficiency; $P$ is the pump power. The SHG efficiency is estimated to be 138\%/W, which is consistent with 155\%/W measured in SHG experiments (Fig.\;\ref{figure2}). The total detection efficiency of 19.4\% is estimated. Therefore, an on-chip squeezing level around 3\;dB can be inferred by setting detection efficiency $T=1$. From linear transmission measurement, we have estimated the optical losses induced by different components. The on-chip waveguide propagation loss is 0.5\;dB. The coupling loss between the thin-film lithium niobate waveguide and output lens is 3.5\;dB. The spatial mode mismatch between LO and squeezed light introduces 0.3\;dB equivalent loss. The photodetector has quantum efficiency 87\%, equivalent to 0.6\;dB loss. The electric noise of the measurement setup is 9.2\;dB lower than the shot noise, equivalent to 0.6\;dB loss \cite{elect_noise_PhysRevA.75.035802}. Therefore, the total loss is 5.5\;dB, corresponding to 28\% detection efficiency. 

\begin{figure}[t]
\centering
\includegraphics[width=\linewidth]{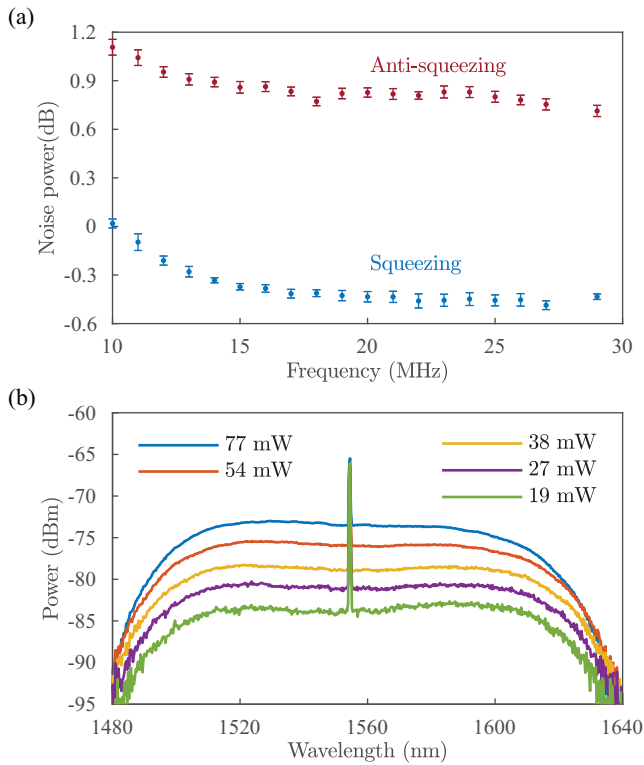}
\caption{Squeezed light bandwidth. (a) Squeezing and anti-squeezing levels as a function of measurement frequency with homodyne detection. (b) Squeezed light spectrum measured by optical spectrum analyzer with different pump powers.}
\label{figure4}
\end{figure}

In order to estimate the bandwidth of squeezed light, we measure the squeezing and anti-squeezing levels at different frequencies (Fig.\;\ref{figure4}(a)). Below 15\;MHz, noise power along both squeezing and anti-squeezing quadrature increases. This is caused by the excess noise from amplified spontaneous emission in EDFA, which is suppressed by the mode-cleaning cavities at high frequencies. Above 30\;MHz, the measurement is limited by the finite bandwidth of the balanced photodetector. In order to measure the full bandwidth of the squeezed light, we directly measure the optical power generated from the parametric down-conversion process at different wavelengths. This is achieved by replacing the homodyne detection setup with an optical spectrum analyzer. Figure\;\ref{figure4}(b) shows the optical spectra at different pump powers. Symmetric spectrum shape is observed with a center wavelength 1554.2\;nm, as idler and signal photons are generated in pairs with constant total energy determined by the pump. The half-width at half maximum is estimated to be above 7\;THz (56\;nm). The generated squeezed light covers the whole C- and L-bands, and half S-band for optical communications. Such large bandwidth will be beneficial to high-speed quantum systems such as quantum communications with wavelength-domain multiplexing and quantum computation with time-domain photonic cluster states \cite{quantum_computing2_Yokoyama2013, roslund2014wavelength}.

Further improvement of the measured squeezing level can be readily achieved by using photodetectors with near-unity quantum efficiency, optimizing spatial overlap visibility, and lowering electronic noise. The squeezing level will be ultimately limited by the coupling efficiency between integrated waveguides and output lens/fiber. Loss below 1.7\;dB has been realized between thin-film lithium niobate waveguides and fiber\;\cite{He:19}. This could lead to a squeezing level above 4.8\;dB, which is sufficient for two-dimensional cluster state generation\;\cite{asavanant2019generation}. The normalized SHG efficiency of our device is $600\%$/W/cm$^2$, which is 10-fold higher than bulk lithium niobate\;\cite{1650535}. Compared with theoretical value $4100\%$/W/cm$^2$, another 7-fold increase can be expected by optimizing the domain inversion process and waveguide thickness uniformity.

In summary, we have demonstrated the generation of squeezed light based on parametric down-conversion using thin-film lithium niobate waveguides. A squeezing level of $0.56\pm0.09$\;dB has been measured with bandwidth up to 7\;THz. This is enabled by the sub-wavelength mode confinement, which leads to a 10-fold increase in pump efficiency. Our sub-wavelength chip-scale platform can enable the integration of squeezed light sources with large-scale photonic circuits for advanced quantum functions. It also enables the realization of second-harmonic generation and fast circuit reconfiguration on the same chip. Therefore, this work will pave the way towards complete on-chip continuous-variable quantum technology.

\begin{backmatter}
\bmsection{Funding} U.S. Department of Energy, Office of Advanced Scientific Computing Research, (Field Work Proposal ERKJ355); Office of Naval Research (N00014-19-1-2190); National Science Foundation (ECCS-1842559, OIA-2040575).

\bmsection{Acknowledgments} We thank Z. Zhang and Y. Xia for helpful discussion about squeezing measurement. 

\bmsection{Disclosures} The authors declare no conflicts of interest.

\bmsection{Data availability} Data may be obtained from the authors upon reasonable request.

\end{backmatter}

\bibliography{bib_intro_quaninfo, bib_intro_chip, bib_LNOI_exp}

\bibliographyfullrefs{bib_intro_quaninfo, bib_intro_chip, bib_LNOI_exp}

\end{document}